# Publish and Perish:

# How AI-Accelerated Writing Without Proportional Verification Investment Degrades Scientific Knowledge

S. Joon Kwon[1]

School of Chemical Engineering, Department of Semiconductor Convergence Engineering,

Department of Future Energy Engineering, Department of Quantum Information Engineering, &

SKKU Institute of Energy Science and Technology (SIEST)

Sungkyunkwan University (SKKU), Suwon 16419, Republic of Korea

## Abstract

Artificial intelligence tools are accelerating manuscript production far faster than peer review capacity can expand. Applying the theory of constraints from manufacturing science, we formalize this asymmetry through a minimal two-variable ordinary differential equation model coupling review queue evolution and verification quality degradation via an endogenous, queue-pressure-driven review AI adoption mechanism. The causal chain is: writing AI adoption increases submissions, growing the review queue, which drives reviewer AI adoption under pressure, degrading verification quality and reducing net knowledge output. Under empirically informed parameters (writing acceleration $\gamma = 2.0$, review acceleration $\delta = 0.5$), the model predicts a deceptive honeymoon where knowledge output peaks at $1.10K_0$ (circa 2026), followed by paradox onset at $t = 6$ years (2028) and long-term degradation to $0.68K_0$ (32% loss), approaching a steady state of $0.60K_0$ (40% loss). The critical condition for net benefit is $\delta > \gamma$; the current operating point lies deep in the paradox regime. Empirical validation against NeurIPS, ICLR, arXiv, and bioRxiv submission data shows qualitative consistency with observed post-ChatGPT acceleration patterns. Policy analysis reveals that only combined interventions such as review infrastructure investment paired with institutional quality standards can restore positive knowledge production.

---

[1] To whom correspondence should be addressed: sjoonkwon@skku.edu, ORCID: 0000-0002-1037-2743

# 1. Introduction

The academic publishing system is experiencing a structural transformation. Major AI conferences report that 10-21% of peer reviews are now AI-generated [1-3], with over half of researchers admitting to using AI during review [4]. Analyses suggest that approximately 14% of recent PubMed abstracts contain AI-generated material [5], and for datasets such as UK Biobank, mass-manufactured papers may now outnumber legitimate analyses by an order of magnitude [6]. Agentic AI systems can produce a complete manuscript, including hypothesis, methods, results, and discussion, in under an hour with minimal human intellectual contribution [7]. Preprint servers document dramatic submission increases across disciplines. The dominant narrative frames these developments as a productivity revolution. We argue instead that they represent a productivity illusion that can lead to systems failure, that the manufacturing sector learned to avoid decades ago.

The theory of constraints (TOC), developed by Goldratt [8], teaches that in any multi-step sequential process, only one step constitutes the bottleneck. Accelerating any non-bottleneck step does not improve total system throughput; rather, it causes excess work-in-progress inventory to accumulate upstream of the bottleneck, increasing lead times and degrading quality. In scientific publishing, the bottleneck has never been manuscript preparation. It has been the depth of verification: peer review, reproducibility checking, and critical evaluation. AI has dramatically accelerated writing while review capacity has barely budged. The predictable consequence is that review queues are exploding, and verification depth is eroding as overloaded reviewers increasingly turn to AI tools that flag formatting errors but miss fundamental flaws [9].

Software engineering provides an instructive analogy. The "move fast and break things" philosophy prioritizes rapid feature delivery over code quality [10-12]. Short-term velocity increases; however, technical debt such as shortcuts, deferred maintenance, and accumulated design compromises also grows silently until the codebase becomes brittle. We observe an analogous dynamic in scientific publishing: AI-accelerated manuscript production accumulates verification debt along with papers published with shallow review, unreproducible methods, and unchecked claims. Each weakly verified paper imposes costs on downstream researchers who must navigate a literature increasingly contaminated with subtle errors [13-15]. Like technical debt, verification debt compounds until the system's capacity to produce reliable knowledge degrades substantially.

While empirical studies have documented rising AI use in manuscript preparation [5, 7] and peer review [1-4], and the peer review crisis has been extensively described [16-18], no prior work provides a quantitative framework linking AI adoption dynamics to knowledge output through coupled feedback mechanisms. We suggest that the present study can fill that gap by constructing a minimal dynamical model with two state variables and a clear one-directional causal chain such that

*writing AI adoption (external, prescribed) → submission surge → queue growth → reviewer AI adoption (endogenous, queue-driven) → quality erosion → knowledge decline.*

The dynamical model enables quantitative identification of the critical condition for net benefit and systematic policy scenario analysis. With empirical validation against real-world data from NeurIPS, ICLR, arXiv, and bioRxiv, the model enables quantitative policy analysis, revealing that the knowledge production crisis is not inevitable but requires deliberate, combined interventions across multiple levers.

## 2. Model Formulation

### 2.1. Causal chain and design philosophy

The model is deliberately minimal: two state variables (review queue Q and verification quality q), one external input (writing AI penetration $\varphi_w(t)$), and one endogenous coupling (review AI penetration $\varphi_r(t)$). The one-directional causal chain can be described as

$$\varphi_w(t)\uparrow \rightarrow S\uparrow \rightarrow Q\uparrow \rightarrow \varphi_r(Q)\uparrow \rightarrow q\downarrow \rightarrow K\downarrow$$

This simple mechanism implies that the queue pressure drives reviewer AI adoption, which then degrades verification quality. Writing AI adoption is treated as an exogenous logistic input fitted to empirical data, rather than modeled with its own differential equations and arbitrary parameters. Review capacity is fixed (reviewers are not enzymes; a longer queue does not produce faster reviewing), and AI provides a modest throughput enhancement. The asymmetry between writing acceleration and review acceleration is the fundamental source of the paradox.

### 2.2. Equations

Writing AI penetration follows a prescribed logistic function:

$$\varphi_w\left(t\right) = \frac{1}{1+\exp\left(-k_w\left(t-t_w\right)\right)}, \ (1)$$

with $k_w = 1.0$ and $t_w = 3.0$ years (corresponding to midpoint at 2025), fitted to survey data indicating greater than 50% researcher adoption by 2025 [4]. The initial point $t = 0$ corresponds to November 2022 (date of ChatGPT release). Review AI penetration is endogenous, driven by queue pressure via a Michaelis–Menten-type saturation:

$$\varphi_r(t) = \frac{Q}{Q + Q_c}, \quad (2)$$

where $Q_c = 2.0$ is the queue level at which 50% of reviewers adopt AI tools. This functional form captures the key behavior: when the queue is small, few reviewers feel pressure to adopt AI; as the queue grows, adoption saturates toward 100%.

The two state variables evolve as:

$$\frac{dQ}{dt} = S_0\left(1 + \gamma\varphi_w(t)\right) - R_{max}\left(1 + \delta\varphi_r(t)\right), \quad (3)$$

$$\frac{dq}{dt} = -\lambda\varphi_r(t)\left(q - q_{min}\right) + \mu\left(1 - \eta\varphi_r(t)\right)\left(1 - q\right). \quad (4)$$

Eq.(3) describes queue evolution where submissions $S(t) = S_0(1 + \gamma\varphi_w)$ grow as writing AI spreads, while review throughput $R(t) = R_{max}(1 + \delta\varphi_r)$ is bounded by reviewer capacity ($R_{max}$) plus a modest AI-assisted enhancement. Critically, R does not depend on Q directly. In other words, reviewer capacity is fixed, which is physically correct for peer review (unlike enzyme kinetics, more substrate does not produce faster processing). Eq.(4) governs verification quality. The first term corresponds to the degradation and pulls quality toward the institutional floor $q_{min}$ at a rate proportional to AI review adoption. The second term corresponds to the restoration and represents human reviewers pulling quality back toward the ideal level (i.e., q = 1). However, this restorative force weakens as AI displaces human review effort (controlled by the displacement fraction $\eta$). The parameter $q_{min}$ represents the institutional minimum standard which is an editorial policy variable enforcing a lower bound on verification depth. Without such guardrails, AI-assisted review could drive quality toward zero at long time.

Knowledge output, K(t), can be simply defined as multiplication K(t) = R(t)q(t), normalized to the pre-AI baseline $K_0 = R_{max} \cdot 1 = 1$. This formulation captures an important insight: K reflects the collective verified knowledge output of the publishing system, not the quality of any individual paper. The linear product form assumes that throughput and quality contribute equally to knowledge output; nonlinear alternatives (i.e., $K = R \times q^\nu$ with the exponent $\nu > 1$ penalizing low quality more heavily) are explored in Section S1 in Supplementary Information (SI) and do not qualitatively change the nature of the paradox.

## 2.3. Parameters

Table 1 summarizes all model parameters with their values and justifications. The model has 9 parameters (excluding logistic shape parameters), of which 2 ($S_0$, $R_{max}$) are normalizations, 2 ($\gamma$, $\delta$) are informed by Amdahl's law and empirical submission data, and 5 ($Q_c$, $\lambda$, $\mu$, $\eta$, $q_{min}$) are explored via sensitivity analysis (Figures 3 and 4).

**Table 1.** Model parameters, values, and justifications

| Symbol | Value | Definition | Justification |
|--------|-------|------------|---------------|
| $\gamma$ | 2.0 | Writing acceleration factor | Amdahl lower bound = 0.3-0.7; empirical NeurIPS CAGR supports = 2.0 (includes community growth) |
| $\delta$ | 0.5 | Review acceleration factor | AI automates approximately 30% of review mechanics leading to net 50% speedup on that fraction |
| $Q_c$ | 2.0 | Queue half-saturation for $\varphi_r$ | Queue level at which 50% of reviewers adopt AI tools |
| $\lambda$ | 0.3 | Quality degradation rate | Rate at which AI review erodes verification depth |
| $\mu$ | 0.5 | Quality restoration rate | Rate at which human review restores quality |
| $\eta$ | 0.8 | Human displacement fraction | AI displaces 80% of human review effort at full adoption |
| $q_{min}$ | 0.2 | Quality floor (policy variable) | Institutional minimum standard; controllable by editors |

Both $S_0$ and $R_{max}$ are normalized to 1.0, representing pre-AI equilibrium where submissions balance reviewer capacity. For writing acceleration ($\gamma = 2.0$), Amdahl's law provides a lower bound on AI-driven pipeline acceleration. Drafting constitutes approximately 30-50% of total research time; LLMs accelerate drafting by 2- or 3-times, yielding a net pipeline speedup of approximately 1.3-1.7-times (i.e., $\gamma = 0.3$-0.7 from drafting alone). However, the observed submission acceleration at AI-intensive venues substantially exceeds this bound: NeurIPS grew at 27% CAGR post-ChatGPT (+128% over year 2020-2025), implying effective $\gamma = 1.5$-2.5. This excess reflects factors beyond drafting speedup. For instance, AI lowered barriers to entry, accelerated literature review and figure generation, and concurrent community expansion. We adopt $\gamma = 2.0$ as an empirically informed central estimate of total submission-rate acceleration in the AI era, acknowledging that it captures both AI-tool effects and correlated secular trends. The model's value lies not in the specific prediction at $\gamma = 2.0$ but in the framework: the critical condition of $\delta > \gamma$ and the policy lever analysis (Figure 4) hold for any $\gamma > \delta$, and the paradox persists across the range $\gamma \in [0.5, 3.0]$ as explored in Figure 3. For review acceleration ($\delta = 0.5$), AI review tools accelerate mechanical tasks (reference verification, plagiarism detection, formatting checks) but do not substitute for deep domain judgment, novelty assessment, or reproducibility verification [3, 4, 18]. Mechanical tasks constitute approximately 30% of total review effort. Even perfect automation of this fraction yields an upper bound of 1/(1 - 0.3) = 1.43 by Amdahl's law. Since the realistic AI contribution to this fraction is partial, $\delta \approx 0.3$-0.5 is a reasonable estimate. We note that even optimistic $\delta$ values do not eliminate the paradox unless $\delta > \gamma$ (Figure 3).

## 2.4. Analytical steady state

At long times ($\varphi_w$ approaching 1), steady-state submissions are $S_{ss} = S_0(1 + \gamma) = 3.0$, while maximum review throughput is $R_{max,eff} = R_{max}(1 + \delta) = 1.5$. Since $S_{ss} > R_{max,eff}$, the queue diverges (Q approaching to ∞), driving $\varphi_r$ toward 1. The steady-state quality is then:

$$q_{ss} = \frac{\lambda q_{min} + \mu(1-\eta)}{\lambda + \mu(1-\eta)} = 0.40, \quad (5)$$

yielding $K_{ss} = R_{max,eff} \times q_{ss} = 1.5 \times 0.40 = 0.60$, or 40% knowledge loss at analytical steady state. The critical condition for the paradox is $S_0(1 + \gamma) > R_{max}(1 + \delta)$, which simplifies (when $S_0 = R_{max}$) to the situation of the paradox occurrence if and only if $\delta < \gamma$. Currently, $\delta = 0.5 \ll \gamma = 2.0$. To avoid the paradox entirely requires $\delta \geq \gamma = 2.0$, meaning review AI must deliver a 200% capacity enhancement which is far beyond current capabilities.

## 3. Results

### 3.1. System dynamics

Figure 1 presents the baseline simulation over a 20-year horizon (initial point t = 0 at November 2022 (ChatGPT release date)). Under baseline parameters ($Q_0 = 0$, $q_0 = 1.0$, representing the pre-AI equilibrium), the model predicts a two-phase trajectory.

The first phase is about deceptive honeymoon (t = 0-6 yr, 2022-2028). In this phase, writing AI adoption spreads rapidly (Figure 1a, blue), with $\varphi_w$ reaching 0.50 by 2025 (t = 3 yr). Review AI adoption (Figure 1a, red) lags substantially, driven not by voluntary adoption but by queue pressure: $\varphi_r$ reaches 0.35 at t = 3 yr and 0.63 at t = 5 yr. Submissions S(t) rise sharply from 1.0 to 2.0 within 3 years (Figure 1b), while review throughput R(t) increases only modestly from 1.0 to 1.18. The widening gap between S and R drives queue accumulation and pressure (Figure 1b, dashed gray). Subsequently, the verification quality is expected to decrease monotonically (Figure 1c). Despite the growing queue, knowledge output K(t) initially increases, peaking at $1.10K_0$ at t = 3.5 yr (circa year 2026; Figure 1d). This honeymoon occurs because early throughput gains (R rises +22%) outpace quality degradation (q falls only to 0.90). All conventional metrics including publications, throughput, turnaround appear healthy while verification debt accumulates invisibly.

In the second phase, the paradox onset and sustained decline (t > 6 yr, post-2028). Knowledge output crosses below $K_0$ at t = 6.1 yr (year 2028), marking paradox onset (Figure 1d, red marker). Thereafter, K declines monotonically as quality erosion overwhelms throughput gains. By t = 10 yr (year 2032), $K/K_0$ = 0.82; by t = 20 yr (year 2042), $K/K_0$ = 0.68 (Figure 1d). Verification quality degrades from q = 1.0 to 0.46 over 20 years (Figure 1c), approaching the analytical steady state value $q_{ss} = 0.40$. The queue grows without bound ($Q(t = 20 \text{ yr}) = 26.8$), confirming the analytical prediction that $S_{ss} > R_{max,eff}$. The implied wait time at $t = 20$ yr is $Q/R = 18$ time units, representing a submission-to-publication pipeline approaching multi-year delays. The decomposition of K decline reveals the mechanism: at $t = 20$ yr, review throughput R has increased by +47% (from 1.0 to 1.47), but quality q has decreased by −54% (from 1.0 to 0.46). The quality loss exceeds the throughput gain, yielding a net 32% knowledge loss. This confirms the core thesis: the problem is quality erosion driven by queue pressure, not throughput insufficiency.

## 3.2. Empirical validation

We compare model predictions against submission data observed from four major venues spanning 2008-2026 (Figure 2). The model predicts submission growth of approximately 200% over 5 years at $\gamma = 2.0$. Empirical data shows: NeurIPS main track grew from 9,467 to 21,575 submissions (+128%, 2020-2025) [19, 20]; ICLR grew from 2,594 to 19,631 (+657%, 2020-2026) [21]; arXiv increased from approximately 15,000 to 24,000 monthly submissions (+60%, 2020-2025; Figure 2c) [22]; and bioRxiv grew from 38,100 to 48,000 annual preprints (+26%, 2020-2025) [23].

      The pre- versus post-ChatGPT submission acceleration pattern is informative. NeurIPS grew at approximately 5% compound annual growth rate (CAGR) during 2020-2022 (pre-ChatGPT), accelerating to approximately 27% CAGR during 2022-2025 (post-ChatGPT, Figure 2a). ICLR exhibits even sharper acceleration: approximately 14% CAGR pre-ChatGPT versus approximately 55% CAGR post-ChatGPT (Figure 2b). arXiv monthly submissions also accelerated, from approximately 8% CAGR pre-ChatGPT to approximately 11% post-ChatGPT (Figure 2c), a more modest increase consistent with its broader disciplinary coverage diluting AI-specific effects. In contrast, bioRxiv (Figure 2d), a biology venue where AI writing tools have lower adoption, shows decelerating growth (from approximately 8% to approximately 3% CAGR), providing a suggestive (though not definitive) comparison. The differential pattern (sharp post-ChatGPT acceleration in AI/CS venues versus deceleration in biology) is consistent with AI writing tools as a contributing factor, though we emphasize that confounding factors (expansion of the AI research community, broadened conference scope, increased industry participation) also contribute to growth at AI/ML venues, and bioRxiv's deceleration may partly reflect saturation of its natural user base after a 50-fold expansion since 2014. A definitive causal test would require within-venue variation in AI tool access, which is not currently available. The model's $\gamma = 2.0$ likely overstates the AI-tool-specific contribution; however, the qualitative prediction of differential acceleration by AI-tool-adoption intensity is supported.

      On the review side, the model predicts $\varphi_r = 0.15$-$0.20$ by year 2-3. Empirical ICLR data shows 15.8% AI-generated reviews in 2024 and approximately 20% in 2025 [1, 2], consistent with model

predictions in order of magnitude. These estimates rely on GPT-text detectors with known false positive and negative rates; the true rates may differ, but the order of magnitude is consistent. Editorial reports from 2024–2025 confirm that manuscripts now routinely take 12 or more months from submission to publication, with 30% of reviewers in high-income countries reporting excessive review requests [16-18].

### 3.3. Parameter space exploration

Figure 3 maps long time behavior of the knowledge measured by $K(t = 20 \text{ yr})/K_0$ across the $(\gamma, \delta)$ parameter space. The $K/K_0 = 1$ contour traces a clear boundary: the system benefits from AI only when review acceleration exceeds writing acceleration (approximately $\delta > \gamma$). Unfortunately, the current observation of the operating point ($\gamma = 2.0$, $\delta = 0.5$) lies deep in the paradox regime. The heatmap in Figure 3 clearly demonstrates that the paradox is robust: it persists across a wide range of parameter combinations, not only at the baseline values. Moderate increases in $\delta$ (i.e., to 1.0 or 1.5) reduce the severity of knowledge degradation but do not eliminate it.

### 3.4. Policy lever analysis

Figure 4 explores two key policy levers: review acceleration $\delta$ (Figure 4a) and quality floor $q_{min}$ (Figure 4b). Varying $\delta$ (Figure 4a) reveals that increasing review acceleration from the baseline $\delta = 0.5$ to $\delta = 1.0$ or 1.5 reduces the severity of knowledge loss but does not eliminate the paradox. Only at $\delta = 2.0$ (the critical threshold) does the system approach long-term stability with $K = K_0$. At $\delta = 2.5$, the system enters the benefit regime with $K > K_0$. Without any review AI enhancement ($\delta = 0$), knowledge loss is most severe. Varying $q_{min}$ (Figure 4b) demonstrates that raising the institutional quality floor (through mandatory code/data availability, enhanced reproducibility checklists, or human-in-the-loop verification for AI-flagged issues) significantly shifts long-term knowledge output upward. At $q_{min} = 0.5$ or higher, the system sustains $K > 0.8K_0$ even under baseline $\delta$. Crucially, $q_{min}$ is directly controllable by editorial policy, making it among the most actionable interventions. This result underscores a key asymmetry: while raising $\delta$ requires technological breakthroughs, raising $q_{min}$ requires only institutional will.

## 4. Discussion

### 4.1. Policy implications

The minimal model identifies two complementary intervention pathways, neither of which is sufficient alone. The first pathway is investing in the bottleneck (increasing $\delta$). The critical condition $\delta > \gamma$ translates to a concrete imperative such that for every unit invested in AI writing tools, proportionally more must be invested in AI-assisted review infrastructure as the tools for reproducibility verification, statistical auditing, and deep methodological checking that amplify human judgment rather than replace it. Currently, the ratio is inverted. Even aggressive review AI investment (i.e., $\delta = 1.5$) fails to eliminate the paradox; only at substantially enhanced review process (i.e., $\delta \geq 2.0$) does the system stabilize. For reviewers, this means that AI-powered tools must progress beyond surface-level checks (formatting, plagiarism detection) toward substantive verification of results and methodology. However, this pathway is limited. Therefore, the second pathway such as raising the quality floor (increasing $q_{min}$) should be considered. Contrary to controlling $\delta$, which requires technological development, $q_{min}$ is directly controllable by editorial and institutional policy. Concrete measures include mandatory code and data availability, reproducibility checklists, pre-registration of analysis plans, human-in-the-loop verification for AI-flagged issues, and post-publication living review. These measures enable continuous annotation, challenge, and correction after publication [24]. Review credit systems that assign DOIs to peer reviews and count them in promotion metrics or credits [25] incentivize thorough reviews, since reviewing is currently uncompensated labor. Open peer review (publishing reviews alongside papers) deters low-effort AI-generated reviews, as several journals (i.e., eLife, EMBO, F1000Research) already demonstrate [26].

For funders and institutions, the model reveals that volume-based quantitative metrics (papers per year, h-index, total citations) become actively misleading in an AI-accelerated regime. When writing is cheap and reviewing is the bottleneck, such metrics reward system-degrading behavior which would lead to a tragedy of the commons in which individually rational manuscript production collectively degrades the knowledge base. Alternative metrics emphasizing verification depth (i.e., reproducibility rates, post-publication correction rates, open data/code compliance) become essential. Capping publications in grant applications, mandating reproducibility documentation [27], and allocating a fraction of research budgets to verification work would directly counter AI-accelerated mass production. These levers are politically feasible because funders are partially insulated from publisher revenue incentives. For researchers individually, the model shows that the honeymoon period is real but brief (approximately 4 years at baseline parameters). Individual rationality is collectively self-defeating: early adopters gain short-term productivity while contributing to the queue pressure that ultimately degrades everyone's knowledge environment.

Editors occupy the most strategically leveraged position. Our model shows that $q_{min}$ is among the most stabilizing parameters (Figure 4b). Raising $q_{min}$ institutionally through mandatory code/data availability, reproducibility checklists, and human verification of AI-flagged issues shifts long-term K upward significantly. Publishers face a political economy problem: revenue scales with volume, not quality. Tracking post-publication correction rates, reproducibility outcomes, and downstream citation patterns as editorial performance metrics would realign incentives with knowledge quality.

Sensitivity analysis (Section S2 in SI) reveals a finding with direct policy relevance: the human displacement fraction $\eta$ is the single most influential parameter (elasticity $-0.97$), exceeding the influence of both $\delta$ (elasticity $+0.35$) and $q_{min}$ (elasticity $+0.29$). This means that controlling how much AI displaces (rather than augments) human review effort is the most effective lever for preserving knowledge quality. Tools designed to amplify reviewer judgment by flagging statistical anomalies, verifying code reproducibility, checking data consistency while keeping humans in the critical evaluation loop, are far more valuable than tools that automate the entire review process. To test human-in-the-loop effects, we did Monte Carlo (MC) simulation (Section S3 in SI) with stochastic noise on submissions and quality. The simulations results confirm that the paradox occurs in 100% of 500 trajectories, with $K(20yr)/K_0 = 0.677 \pm 0.032$, closely matching the deterministic prediction of 0.676 and demonstrating robustness of the dynamics to (possibly human-related) stochastic perturbations.

## 4.2. Limitations

We should also mention several limitations which would merit discussion and future studies. First, the model assumes homogeneous researcher and reviewer populations. In reality, although tested in MC simulations, AI adoption varies by discipline (physics versus biology), career stage (junior versus senior), geography or ethnicity, and institutional context; a disaggregated model stratified by field could reveal discipline-specific vulnerabilities. Second, the scalar value $q$ aggregates multi-aspect review (methodological rigor, novelty assessment, statistical validity, ethical review) that may degrade at different rates under AI pressure. Third, the model assumes passive parameter evolution, not capturing strategic gaming dynamics. For example, authors deliberately overloading competitors' review queues, reviewers weaponizing AI to reject rivals, or publishers exploiting AI to maximize volume at quality's expense. Fourth, submission growth at AI/ML venues reflects multiple concurrent trends beyond AI writing tools: the AI research community has expanded enormously since 2020, industry R&D investment has surged, and conferences have broadened their scope (NeurIPS added datasets and position paper tracks). Our model treats submission growth as driven primarily by writing acceleration, which may overstate AI tools' contribution. However, the differential acceleration pattern (AI-intensive venues accelerating sharply post-ChatGPT while biology venues decelerate) supports the model's qualitative prediction even if the quantitative attribution is approximate. Fifth, the model does not include editorial desk rejection as a separate control variable; incorporating queue-aware desk rejection (reducing the fraction of submissions that enter peer review when the queue exceeds a threshold) would allow more nuanced policy analysis and is a natural extension. Sixth, and importantly, the verification quality trajectory $q(t)$ is the model's central prediction but lacks direct empirical validation: while submission growth $S(t)$ is validated against venue data and reviewer AI adoption $\varphi_r$ is validated against Liang *et al.* [1, 2], no comparable time-series data exists for aggregate review depth. Future empirical work tracking retraction rates, post-publication correction frequencies, review length and substantiveness, or reproducibility metrics over time would provide the most valuable test of the model's core prediction.

Despite these limitations, the structural prediction that asymmetric acceleration degrades net knowledge output is robust across wide ranges of the parameters (Figure 3 and Section S2 in SI). Latin Hypercube Sampling across all seven model parameters shows the paradox occurring in 64% of 500 random parameter combinations, rising to 100% when parameters are constrained to empirically plausible ranges (Section S2 in SI). These results suggest the paradox is a genuine systemic vulnerability rather than a calibration artifact.

## 5. Conclusion

Goldratt's theory of constraints demonstrates that optimizing a non-bottleneck is not merely wasteful but destructive. Our minimal model shows that AI-accelerated writing, unmatched by proportional investment in review infrastructure, produces a paradoxical decrease in verified knowledge output through a clear causal mechanism: queue pressure drives reviewer AI adoption, which degrades verification quality faster than throughput increases. The system exhibits a deceptive honeymoon period (peaking at $1.10K_0$ in 2026) before declining to $0.68K_0$ at 20 years (32% knowledge loss) and approaching an analytical steady state of $0.60K_0$ (40% loss). The critical condition is $\delta > \gamma$: review acceleration must exceed writing acceleration. Currently, $\gamma = 2.0$ while $\delta = 0.5$, placing academic publishing firmly in the paradox regime leading to knowledge degradation. The paradox is not inevitable. Escaping it requires investing in the bottleneck: review infrastructure that amplifies human judgment (increasing $\delta$), institutional quality standards that raise the minimum quality floor $q_{min}$, and metric reform that values verification depth over publication volume. Neither technology investment nor regulation alone suffices; only the combination breaks the feedback loop linking queue pressure to quality erosion. The counterintuitive lesson is clear: in science, faster writing without proportional verification investment creates less knowledge, not more. The choice facing the research community is whether to invest deliberately in the bottleneck now, or to accept a future where we publish prolifically while knowing less.

## Methods

**Numerical methods.** Integration was performed using the Runge–Kutta method (RK45, scipy.integrate.solve_ivp) over t $\in$ [0, 20] years with maximum time step 0.02, relative tolerance $10^{-10}$, and absolute tolerance $10^{-12}$. The time axis is discretized into 2,000 evaluation points. Parameter space exploration (Figure 3): each pixel represents an independent 20-year simulation; the $K/K_0 = 1$ contour is computed via Matplotlib's contour routine with grid resolution 60×60. Policy sweeps (Figure 4): each curve represents an independent simulation with a single parameter varied. Initial conditions: $Q_0 = 0$ (no pre-AI backlog), $q_0 = 1.0$ (ideal pre-AI quality).

**Empirical data sources.** NeurIPS submissions: 2008-2025 main track data from NeurIPS blog [19] and Paper Copilot [20]. ICLR submissions: 2013-2026 data from Paper Copilot and OpenReview [21]. arXiv submissions: monthly rates from arxiv.org [22]. bioRxiv preprints: annual totals [23]. ICLR AI review detection rates from [1, 2]. Nature AI usage survey from [4]. Peer review crisis reports from [16-18].

**Code availability.** Model implementation: Python 3.12, NumPy 1.26, SciPy 1.11, Matplotlib 3.8. Full code available at https://github.com/sjoonkwon0531/publish-and-perish.

## Author Contributions

S.J.K.: Conceptualization, Methodology, Software, Formal analysis, Writing – original draft, Writing – review & editing, Visualization

## Competing interests

The author declares no competing interests.

## Acknowledgments

This work was partially supported by the Samsung Research Funding Center for Samsung Electronics under project number SRFC-MA2201-02.

## Figure Captions

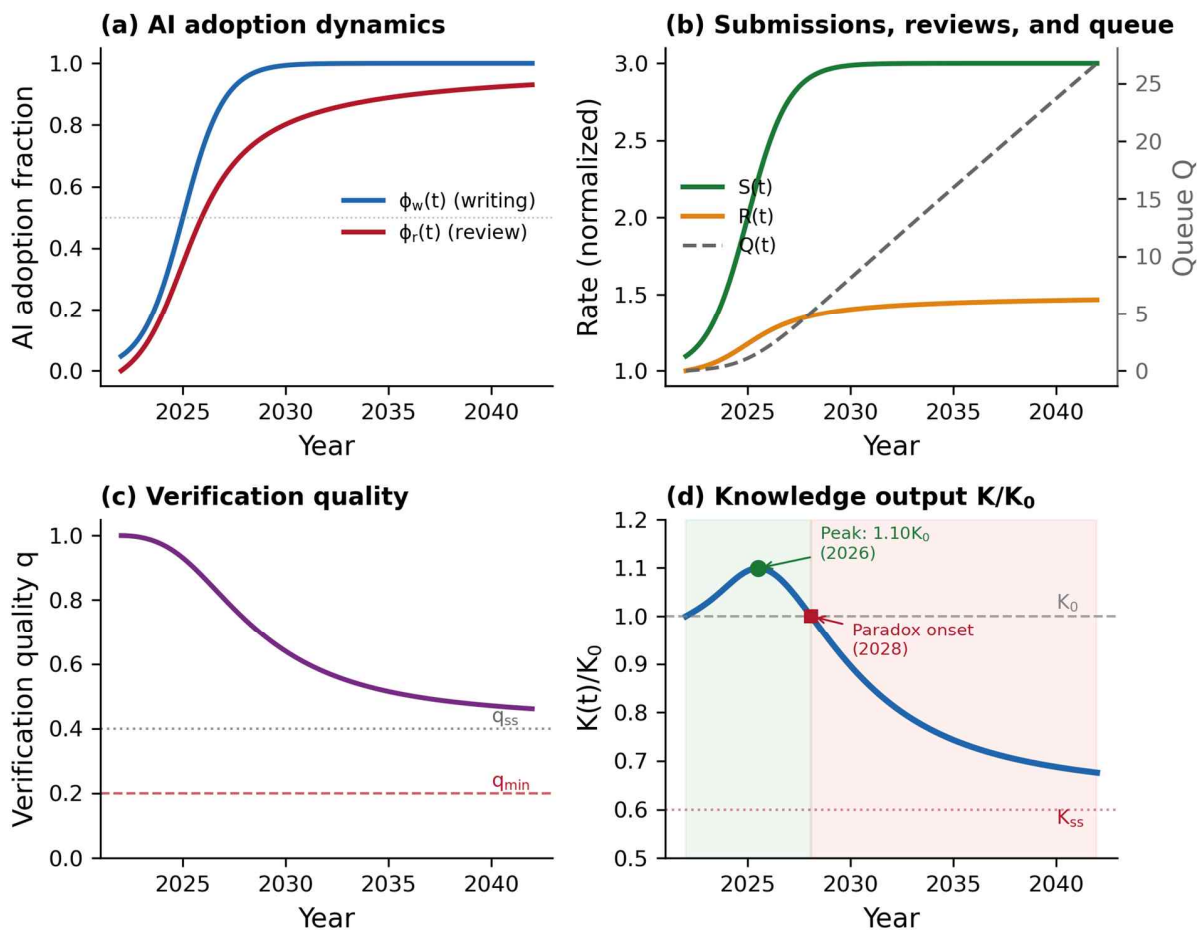

**Figure 1.** Model dynamics of the Publish and Perish paradox. (a) AI adoption in writing ($\varphi_w$, external logistic) and in review ($\varphi_r$, endogenous, queue-driven). (b) Submission rate $S(t)$, review throughput $R(t)$, and review queue $Q(t)$ (right axis, dashed). (c) Verification quality $q(t)$, with analytical steady state $q_{ss} = 0.40$ and quality floor $q_{min} = 0.20$ marked. (d) Normalized knowledge output $K(t)/K_0$, showing honeymoon peak at $1.10K_0$ (2026) and paradox onset at $t \approx 6$ yr (2028). Analytical $K_{ss}/K_0 = 0.60$ shown as lower bound.

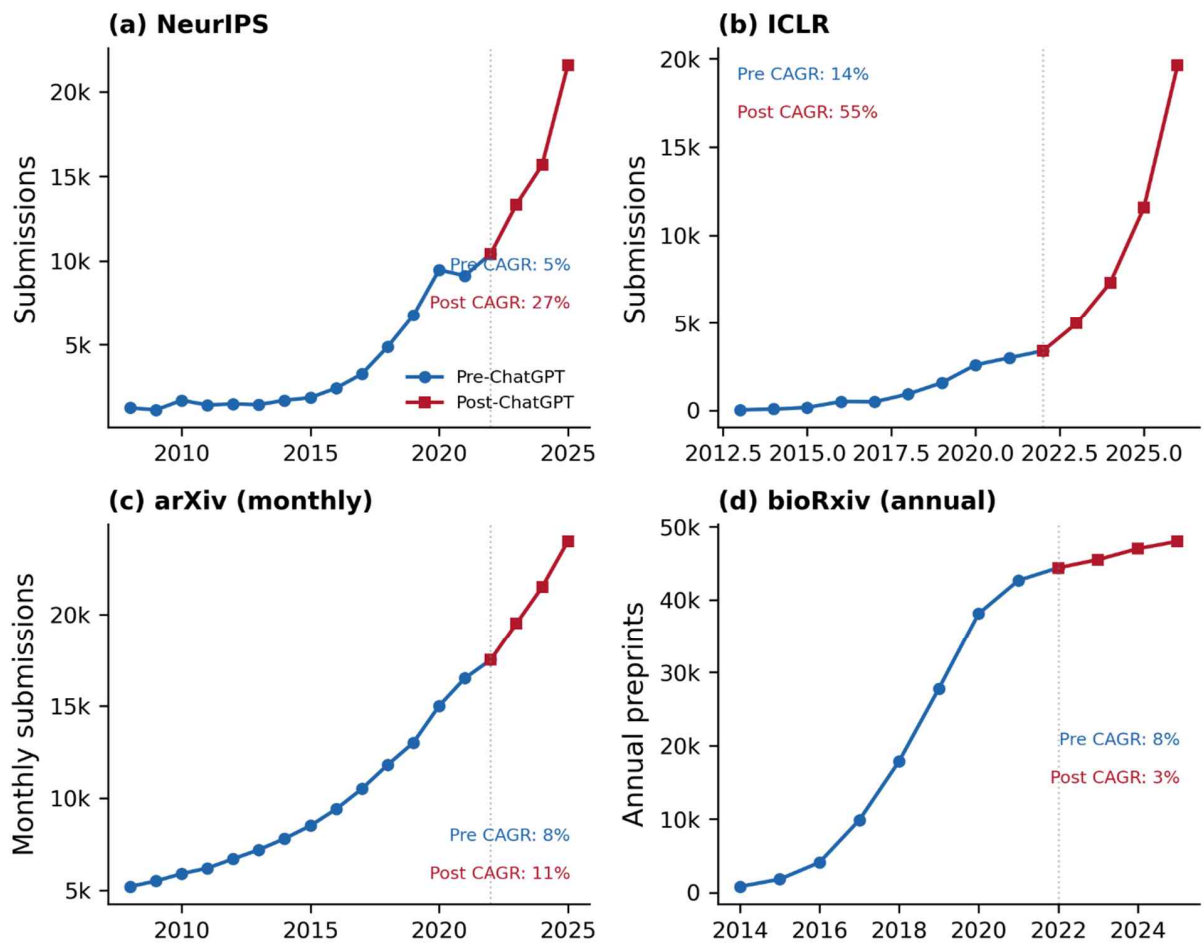

**Figure 2.** Empirical validation. Annual submissions for (a) NeurIPS (2008-2025), (b) ICLR (2013-2026), (c) arXiv monthly (2008-2025), and (d) bioRxiv annual (2014-2025). Circles: pre-ChatGPT data; squares: post-ChatGPT data. CAGR shown for each period. AI-intensive venues (a-c) accelerated post-ChatGPT; bioRxiv (d) decelerated, providing a suggestive comparison. Data sources: [19-23].

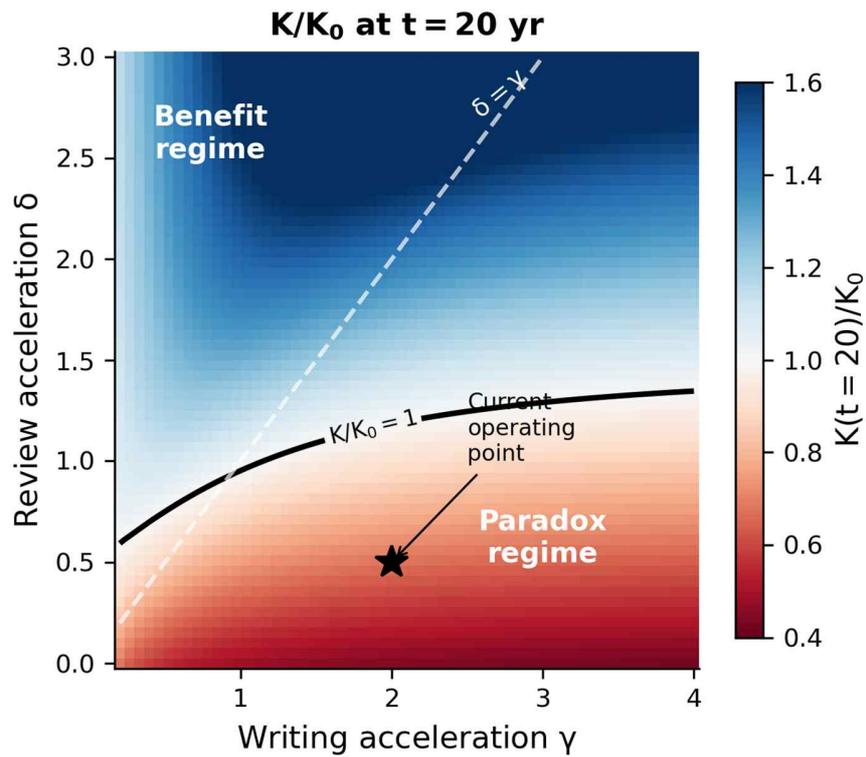

**Figure 3.** Parameter space exploration of the knowledge degradation. Heatmap of K(t = 20)/K₀ as a function of writing acceleration γ and review acceleration δ. The K/K₀ = 1 contour (black) separates the paradox regime (red, below) from the benefit regime (blue, above). The dashed white line marks δ = γ (simplified critical condition). Black star: current observed operating point (γ = 2.0, δ = 0.5).

... 

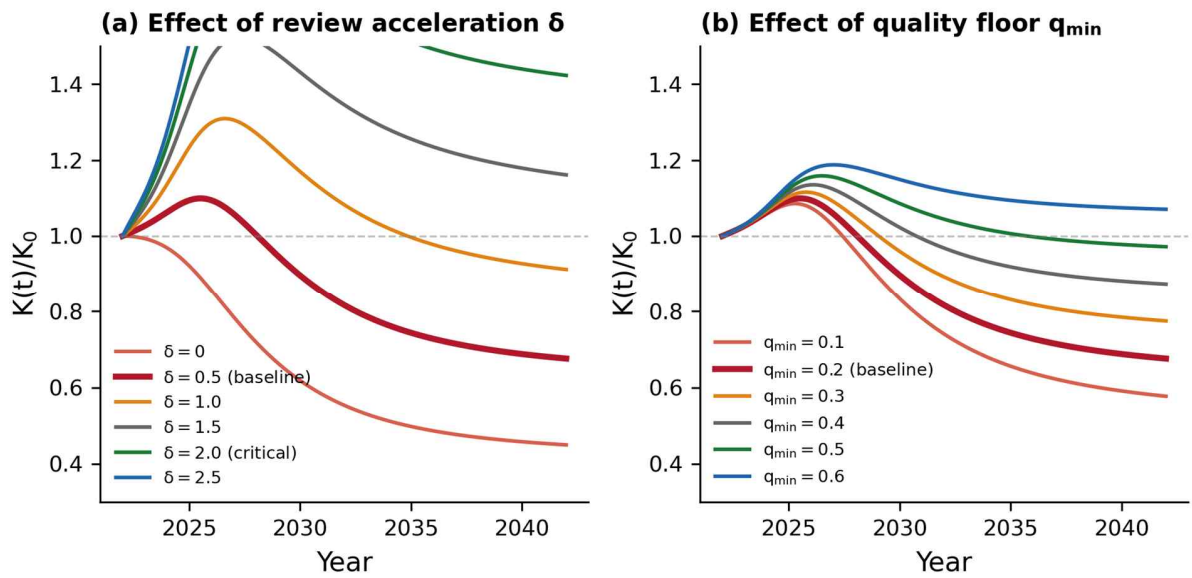

**Figure 4.** Policy lever analysis. (a) Effect of review acceleration δ on K(t)/K₀ over 20 years (year 2022-2042). Baseline δ = 0.5 (bold red); critical threshold δ = 2.0 (green) restores K ≈ K₀. (b) Effect of quality floor $q_{min}$ on K(t)/K₀. Higher institutional standards ($q_{min}$ = 0.4-0.6) significantly mitigate knowledge loss.